# Success Rate and Entanglement Evolution in Search Algorithm

Arti Chamoli and C. M. Bhandari
Indian Institute of Information Technology Allahabad, Deoghat, Jhalwa,
Allahabad - 211011, India.
Email: achamoli@iiita.ac.in, cmbhandari@yahoo.com

Evolution of entanglement with the processing of quantum algorithms affects the outcome of the algorithm. Particularly, the performance of Grover's search algorithm gets worsened if the initial state of the algorithm is an entangled one. Biham et al [14] have shown that the success probability of search algorithm can be seen as an operational measure of entanglement. Following the same line of thought, analytical expressions for entanglement measure for three and five qubit systems have been derived, which reveals that entanglement measure based on Grover's search algorithm is valid for any state with real coefficients for even number of qubits.

**Introduction**

Quantum entanglement [1] and superposition [2] are the pillars of quantum computation and quantum information theory [3, 4]. Quantum information theory has reached the new arenas exploiting these two. Quantum entanglement, inherently a non-classical phenomenon, signifies correlations between quantum systems even if they are space-like separated. In recent times it has been reckoned as a physical resource and hence utilized for various computational tasks including quantum information processing [5] and cryptography [6]. Applications like quantum teleportation can only be materialized if certain amount of entanglement exists between the communicators initially. For this reason, quantification of entanglement of quantum states attains utmost importance. Various entanglement evaluating measures have been figured by current researchers. Methodology based on operational considerations has been successfully employed to formulate entanglement measures for bipartite systems [7, 8]. Based on correspondences between thermodynamics and entanglement, entropy of entanglement has been considered as the unique measure of entanglement of pure states [9]. In addition to this, certain attributes have been framed to measure entanglement [10-13]. These attributes are based on axiomatic considerations. According to these, any entanglement should not prevail in product states, it should not vary with local unitary operations and should not increase consequential to any sequence of local operations complemented by only classical communication between parties. Measures satisfying the above properties are called entanglement monotones [11].

Recently an entanglement measure has been developed by Biham et al [14]. The measure is based on the linkage of the success of Grover's search algorithm [15, 16] to the amount of entanglement present in the initial state. Performance of Grover's algorithm deteriorates with increasing entanglement in the initial state. Considering the modified quantum search as given in [14] in which a product of arbitrary local operations is applied to initial input register, the formulation of maximal probability of success, $P_{max}(\Psi)$, as an entanglement monotone can be precisely made. For a search space containing $N=2^n$ elements, where n is an integer, the elements can be represented by an n-qubit register and the initial register as $|\Phi\rangle$. For a single marked solution, $s$, to the search problem, $P_{max}$ in terms of the operator $U_G^m$, representing $m$ Grover iterations may be written as

$$P_{max} = \max_{U_1,......U_n} \frac{1}{N} \sum_{s=0}^{N-1} \left|\langle s|U_G^m(U_1 \otimes U_2 \otimes ....U_n)|\Phi\rangle\right|^2$$

(1)

by averaging uniformly over all $N$ possible values for $s$. The maximization done is over all local unitary operations $U_1,......U_n$ on the respective qubits of input register state $|\Phi\rangle$. This can be generalized by considering the action of the Grover iterations on the uniform superposition state $|\eta\rangle = \frac{1}{\sqrt{N}} \sum_x |x\rangle$. Applying Grover operator yields,

$$U_G^m|\eta\rangle = |s\rangle + O\left(\frac{1}{\sqrt{N}}\right)$$

(2)

where the second term is a small correction because Grover's algorithm yields a solution with probability $1 - O\left(\frac{1}{\sqrt{N}}\right)$. Multiplying eq. (2) by $(U_G^m)^\dagger$ and then taking the Hermitian Conjugate gives

$$\langle s|U_G^m = \langle \eta| + O\left(\frac{1}{\sqrt{N}}\right)$$

(3)

Substituting in eq. (1) gives, for a general state $|\Phi\rangle$,

$$P_{max} = \max_{U_1,......U_n} \frac{1}{N} \sum_{s=0}^{N-1} \left|\langle \eta|U_G^m(U_1 \otimes U_2 \otimes ....U_n)|\Phi\rangle\right|^2 + O\left(\frac{1}{\sqrt{N}}\right)$$

(4)

Since $|\eta\rangle$ is a product state, eq(4) may equivalently may be expressed as

$$P_{max} = \max_{|e_1,....e_n\rangle} \left|\langle e_1,....e_n|\Phi\rangle\right|^2 + O\left(\frac{1}{\sqrt{N}}\right)$$

(5)

where the maximization now runs over all product states, $|e_1,....e_n\rangle = |e_1\rangle \otimes ...... \otimes |e_n\rangle$, of the $n$ qubits. This suggests that $P_{max}$ depends on the maximum of the overlap between all product states and the input state $|\Phi\rangle$. For a product state as input state, $P_{max}$ would

be equal to one, whereas with an entangled state as input state, $P_{max}$ would never be one. Success probability of the search algorithm depends on the entanglement of initial register state. Quantifying entanglement following the above approach is related to the performance of the quantum state as an input to the modified search algorithm. The measure thusly referred to as Groverian entanglement can be defined for a state $|\Psi\rangle$ by

$$G(\Psi) = \sqrt{1 - P_{max}}$$

(6)

$P_{max}(\Psi)$ is an entanglement monotone and consequently $G(\Psi)$ too. In this letter, the authors have evaluated the success rate of Grover's search algorithm for three and five qubit states and Groverian entanglement measure has been formulated for the same.

**Three-qubit states**

An arbitrary initial state $|\Psi\rangle$ of three qubits can be written as

$$|\Psi\rangle = \sum_{i=0}^{7} a_i |i\rangle$$

(7)

where $|i\rangle = |i_0 i_1 i_2\rangle$.

Following the steps given by Biham et al. [14], an analytical expression for maximum success probability of Grover's search algorithm for three qubit states can be written as

$$P_{max} = \frac{1}{16}\left[\sqrt{(a_0 - a_6 - a_5 - a_3)^2 + (a_4 + a_2 + a_1 - a_7)^2} + \sqrt{(a_0 - a_6 + a_5 + a_3)^2 + (a_4 + a_2 - a_1 + a_7)^2} \right.$$
$$\left. + \sqrt{(a_0 + a_6 - a_5 + a_3)^2 + (a_4 - a_2 + a_1 + a_7)^2} + \sqrt{(a_0 + a_6 + a_5 - a_3)^2 + (a_4 - a_2 - a_1 - a_7)^2}\right]^2$$

(8)

With the analytical expression for, $P_{max}(\Psi)$, for three qubit system, $G(\Psi)$, can be calculated for various choices of $|\Psi\rangle$ with real coefficients.

For a product state, the entanglement measure, as expected, must vanish. This can be easily verified for a uniform product state can be obtained with three qubits in either of these states:

$$\frac{1}{\sqrt{2}}(|0\rangle + |1\rangle), \frac{1}{\sqrt{2}}(|0\rangle - |1\rangle)$$

The substitution of $a_i's$ in the analytical expression gives $P_{max}(\Psi) = 1$, thereby giving $G(\Psi) = 0$.

As given by Biham et al [14], for a maximally entangled GHZ state, $G(\Psi)$ should be $(1/\sqrt{2})$, independent of the number of qubits. However this does not hold correct on substituting the values of coefficients of the state, $\frac{1}{\sqrt{2}}(|000\rangle + |111\rangle)$, in eq.(8). Eq. rather gives $G(\Psi) = 0$ for any three qubit GHZ state which is contrary to the property of maximally entangled states.

For W-states, $P_{max}(\Psi_W) = \left(1 - \frac{1}{n}\right)^{n-1}$ [14]. Hence for three qubit W-state $P_{max}$ must be around 0.44. On substituting the coefficients of the state, $\frac{1}{\sqrt{3}}(|001\rangle + |010\rangle + |100\rangle)$, in eq.(8) gives $P_{max} = 0.75$.

**Five Qubits**
Following the same procedure as earlier, an analytical expression for the maximum success probability of Grover's search algorithm for five qubit states has been obtained from a five-qubit pure state of the form

$$|\Psi\rangle = \sum_{i=0}^{31} a_i |i\rangle, \text{ where } |i\rangle = |i_0....i_4\rangle$$

A general five qubits state is a linear combination of thirty two terms. Thus maximum success probability and hence Groverian entanglement can be obtained for five qubit states with real coefficients from the analytical expression. The expression for the maximum success probability is given in the Appendix.

For example, the general product state with uniform superposition is of the form $\sum_{i=0}^{31} a_i |i\rangle$, where $|i\rangle = |i_0....i_4\rangle$ with $a_i = \frac{1}{4\sqrt{2}}$ for all i. On substituting the value of coefficients of each state it can be seen that $P_{max}(\Psi) = 1$, and thus $G(\Psi) = 0$, verifying that a product state has zero entanglement. Similar to three qubit system, the value of $P_{max}(\Psi)$ for five qubit maximally entangled- and W-states, is not in agreement with the results reprted by Biham et al. [14].

$P_{\max}(\Psi)$ for a maximally entangled state (GHZ state) of the type, $\frac{1}{\sqrt{2}}(|00000\rangle + |11111\rangle)$, comes out to be 1 on substituting the coefficients of the given state in the expression given in appendix. This again contradicts statements for GHZ states [14]. For five qubit W-states as well, $P_{\max}(\Psi) = 0.7$, whereas the value reported by BIham et al [14] is around 0.41.

**Discussion and conclusion**
Following the approach given in [14], authors (17) have examined the success rate of Grover's search algorithm for various four qubit states and Groverian entanglement measure has been worked out for certain kind of input states. For four qubit states with real coefficients, the results, for GHZ- and W-states as well, are in agreement with the values reported in [14]. Biham et al. [14] have derived an explicit analytical expression for two-qubit states. The results for *n* = 2, 3, 4, and 5-qubit states show that entanglement measure based on success probability of Grover's search algorithm is an efficient measure for any arbitrary state with real coefficients for *n* = 2 and 4-qubit systems. Whereas for *n* = 3 and 5-qubit states, the analytical expression is correct only for certain states with real coefficients. Observing the expressions for $P_{\max}(\Psi)$ for two to five-qubit systems, the expression for higher no. of qubits can be anticipated. Thus it can be safely deduced that this operational definition of entanglement measure based on the success rate in attaining the desired state in Grover's algorithm is valid for quantum systems with even number of qubits with real coefficients. For quantum systems with odd number of qubits this is not always true.

**Acknowledgement**
Authors are thankful to Dr. M. D. Tiwari for his keen interest and support. Arti Chamoli is thankful to IIIT, Allahabad for financial support.

## APPENDIX

$$P_{max}(\Psi) = \frac{1}{256}[\sqrt{\{(a_0 - a_{24} - a_{20} - a_{12} - a_{18} - a_{10} - a_6 + a_{30} - a_{17} - a_9 - a_5 + a_{29} - a_3 + a_{27} + a_{23} + a_{15})^2 + (a_{16} + a_8 + a_4 - a_{28} + a_2 - a_{26} - a_{22} - a_{14} + a_1 - a_{25} - a_{21} - a_{13} - a_{19} - a_{11} - a_7 + a_{31})^2\}}$$

$$+\sqrt{\{(a_0 - a_{24} - a_{20} - a_{12} - a_{18} - a_{10} - a_6 + a_{30} + a_{17} + a_9 + a_5 - a_{29} + a_3 - a_{27} - a_{23} - a_{15})^2 + (a_{16} + a_8 + a_4 - a_{28} + a_2 - a_{26} - a_{22} - a_{14} - a_1 + a_{25} + a_{21} + a_{13} + a_{19} + a_{11} + a_7 - a_{31})^2\}}$$

$$+\sqrt{\{(a_0 - a_{24} - a_{20} - a_{12} + a_{18} + a_{10} + a_6 - a_{30} - a_{17} - a_9 - a_5 + a_{29} + a_3 - a_{27} - a_{23} - a_{15})^2 + (a_{16} + a_8 + a_4 - a_{28} - a_2 + a_{26} + a_{22} + a_{14} + a_1 - a_{25} - a_{21} - a_{13} + a_{19} + a_{11} + a_7 - a_{31})^2\}}$$

$$+\sqrt{\{(a_0 - a_{24} - a_{20} - a_{12} + a_{18} + a_{10} + a_6 - a_{30} + a_{17} + a_9 + a_5 - a_{29} - a_3 + a_{27} + a_{23} + a_{15})^2 + (a_{16} + a_8 + a_4 - a_{28} - a_2 + a_{26} + a_{22} + a_{14} - a_1 + a_{25} + a_{21} + a_{13} - a_{19} - a_{11} - a_7 + a_{31})^2\}}$$

$$+\sqrt{\{(a_0 - a_{24} + a_{20} + a_{12} - a_{18} - a_{10} + a_6 - a_{30} - a_{17} - a_9 + a_5 - a_{29} - a_3 + a_{27} - a_{23} - a_{15})^2 + (a_{16} + a_8 - a_4 + a_{28} + a_2 - a_{26} + a_{22} + a_{14} + a_1 - a_{25} + a_{21} + a_{13} - a_{19} - a_{11} + a_7 - a_{31})^2\}}$$

$$+\sqrt{\{~(a_0 - a_{24} + a_{20} + a_{12} - a_{18} - a_{10} + a_6 - a_{30} + a_{17} + a_9 - a_5 + a_{29} + a_3 - a_{27} + a_{23} + a_{15}~)^2 + (a_{16} + a_8 - a_4 + a_{28} + a_2 - a_{26} + a_{22} + a_{14} - a_1 + a_{25} - a_{21} - a_{13} + a_{19} + a_{11} - a_7 + a_{31})^2~\}}$$

$$+\sqrt{\{~(a_0 - a_{24} + a_{20} + a_{12} + a_{18} + a_{10} - a_6 + a_{30} - a_{17} - a_9 + a_5 - a_{29} + a_3 - a_{27} + a_{23} + a_{15}~)^2 + (a_{16} + a_8 - a_4 + a_{28} - a_2 + a_{26} - a_{22} - a_{14} + a_1 - a_{25} + a_{21} + a_{13} + a_{19} + a_{11} - a_7 + a_{31})^2~\}}$$

$$+\sqrt{\{~(a_0 - a_{24} + a_{20} + a_{12} + a_{18} + a_{10} - a_6 + a_{30} + a_{17} + a_9 - a_5 + a_{29} - a_3 + a_{27} - a_{23} - a_{15}~)^2 + (a_{16} + a_8 - a_4 + a_{28} - a_2 + a_{26} - a_{22} - a_{14} - a_1 + a_{25} - a_{21} - a_{13} - a_{19} - a_{11} + a_7 - a_{31})^2~\}}$$

$$+\sqrt{\{~(a_0 + a_{24} - a_{20} + a_{12} - a_{18} + a_{10} - a_6 - a_{30} - a_{17} + a_9 - a_5 - a_{29} - a_3 - a_{27} + a_{23} - a_{15}~)^2 + (a_{16} - a_8 + a_4 + a_{28} + a_2 + a_{26} - a_{22} + a_{14} + a_1 + a_{25} - a_{21} + a_{13} - a_{19} + a_{11} - a_7 - a_{31})^2~\}}$$

$$+\sqrt{\{~(a_0 + a_{24} - a_{20} + a_{12} - a_{18} + a_{10} - a_6 - a_{30} + a_{17} - a_9 + a_5 + a_{29} + a_3 + a_{27} - a_{23} + a_{15}~)^2 + (a_{16} - a_8 + a_4 + a_{28} + a_2 + a_{26} - a_{22} + a_{14} - a_1 - a_{25} + a_{21} - a_{13} + a_{19} - a_{11} + a_7 + a_{31})^2~\}}$$

$$+\sqrt{\{~(a_0 + a_{24} - a_{20} + a_{12} + a_{18} - a_{10} + a_6 + a_{30} - a_{17} + a_9 - a_5 - a_{29} + a_3 + a_{27} - a_{23} + a_{15}~)^2 + (a_{16} - a_8 + a_4 + a_{28} - a_2 - a_{26} + a_{22} - a_{14} + a_1 + a_{25} - a_{21} + a_{13} + a_{19} - a_{11} + a_7 + a_{31})^2~\}}$$

$$+\sqrt{\{~(a_0 + a_{24} - a_{20} + a_{12} + a_{18} - a_{10} + a_6 + a_{30} + a_{17} - a_9 + a_5 + a_{29} - a_3 - a_{27} + a_{23} - a_{15}~)^2 + (a_{16} - a_8 + a_4 + a_{28} - a_2 - a_{26} + a_{22} - a_{14} - a_1 - a_{25} + a_{21} - a_{13} - a_{19} + a_{11} - a_7 - a_{31})^2~\}}$$

$$+\sqrt{\{~(a_0 + a_{24} + a_{20} - a_{12} - a_{18} + a_{10} + a_6 + a_{30} - a_{17} + a_9 + a_5 + a_{29} - a_3 - a_{27} - a_{23} + a_{15}~)^2 + (a_{16} - a_8 - a_4 - a_{28} + a_2 + a_{26} + a_{22} - a_{14} + a_1 + a_{25} + a_{21} - a_{13} - a_{19} + a_{11} + a_7 + a_{31})^2~\}}$$

$$+\sqrt{\{~(a_0 + a_{24} + a_{20} - a_{12} - a_{18} + a_{10} + a_6 + a_{30} + a_{17} - a_9 - a_5 - a_{29} + a_3 + a_{27} + a_{23} - a_{15}~)^2 + (a_{16} - a_8 - a_4 - a_{28} + a_2 + a_{26} + a_{22} - a_{14} - a_1 - a_{25} - a_{21} + a_{13} + a_{19} - a_{11} - a_7 - a_{31})^2~\}}$$

$$+\sqrt{\{~(a_0 + a_{24} + a_{20} - a_{12} + a_{18} - a_{10} - a_6 - a_{30} - a_{17} + a_9 + a_5 + a_{29} + a_3 + a_{27} + a_{23} - a_{15}~)^2 + (a_{16} - a_8 - a_4 - a_{28} - a_2 - a_{26} - a_{22} + a_{14} + a_1 + a_{25} + a_{21} - a_{13} + a_{19} - a_{11} - a_7 - a_{31})^2~\}}$$

$$+\sqrt{\{}(a_0 + a_{24} + a_{20} - a_{12} + a_{18} - a_{10} - a_6 - a_{30} + a_{17} - a_{25} - a_{21} - a_{29} - a_3 - a_{27} - a_{23} + a_{15})^2 + (a_{16} - a_8 - a_4 - a_{28} - a_2 - a_{26} - a_{22} + a_{14} - a_1 - a_{25} - a_{21} + a_{13} - a_{19} + a_{11} + a_7 + a_{31})^2 \;\}\Big]^2$$